# Thermodynamic Fidelity of Generative Models for Ising System

Brian H. Lee, Kat Nykiel, Ava E. Hallberg, Brice Rider, Alejandro Strachan[a]

Author affiliations
*School of Materials Engineering and Birck Nanotechnology Center, Purdue University, West Lafayette, Indiana 47907, USA*

Author email
[a] Author to whom correspondence should be addressed: strachan@purdue.edu

## Abstract

Machine learning has become a central technique for modeling in science and engineering, either complementing or as surrogates to physics-based models. Significant efforts have recently been devoted to models capable of predicting field quantities but the limitations of current state-of-the-art models in describing complex physics are not well understood. We characterize the ability of generative diffusion models and generative adversarial networks (GAN) to describe the Ising model. We find diffusion models trained using equilibrium configurations obtained using Metropolis Monte Carlo for a range of temperatures around the critical temperature can capture average thermodynamic variables across the phase transformation and extrapolate to higher and lower temperatures. The model also captures the overall trends of physical properties associated with fluctuations (specific heat and susceptibility) except at the non-ergodic low temperatures and non-trivial scale-free correlations at the critical temperature, albeit with some difference in the critical exponent compared to Monte Carlo simulations. GANs perform more poorly on thermodynamic properties and are susceptible to mode-collapse without careful training. This investigation highlights the potential and limitations of generative models in capturing the complex phenomena associated with certain physical systems.

# 1. Introduction

Physics-based simulations are indispensable tools for scientific discovery and engineering. For example, molecular dynamics (MD) and Monte Carlo (MC) simulations are central in biology,[1] material science,[2,3] and chemistry.[4,5] Over the last decades, machine learning (ML) models have been successfully used to enhance such models and to learn from them. Examples include machine-learning-based interatomic potentials,[6–8] multiscale modeling,[9,10] mapping of microstructure to material properties,[11,12] extracting interpretable physical laws,[13,14] and prediction of molecular structural properties.[15,16] These successes and many related ones have significantly accelerated progress in the field and have motivated ongoing investigations to extend the use of ML methods to a wider range of scientific tasks.

Generative models, including generative adversarial network (GAN) and diffusion models,[17–19] have gathered significant interest since their introduction and demonstrated remarkable prowess in synthesizing realistic data samples, including images, text, and videos. In addition, these models have been utilized for scientific investigations, such as the generation of molecular conformations,[20–23] and materials' microstructures,[24] as well as calorimeter simulations.[25] These studies have shown that the generative models can produce data required for scientific investigations at orders of magnitude improved computational efficiency compared to traditional simulations. Given the growing use of these models to represent physical systems it is imperative to establish their range of applicability.

In this study, we evaluate the ability of diffusion models and GAN to generate configurations of the Ising model which describes the interaction of spins and captures the non-trivial physics involved in first- and second-order phase transitions between a non-magnetic and a ferromagnetic state.[26] Due to its simplicity and the existence of an analytical solution[27] for the non-trivial 2D problem, Ising has played a central role in studies of critical phenomena[28–30] and has been applied or extended to a wide range of fields, from biological processes[31–33] to crystal plasticity.[34] Recent papers[35–37] have applied GAN and diffusion models to model Ising and several studies[38–40] have investigated the performance of other generative models. However, ability of GAN and diffusion to capture the subtle physics of the model near the critical point as well as the diversity of generated ensembles and the equilibration process have not been well examined. Our analyses demonstrate that the diffusion model can generate configurations that result in accurate thermodynamic properties when trained with configurations at various temperatures across the phase transition. Achieving similar accuracy with GAN models requires adding the thermodynamic quantities of interest to the objective function during training. While the diffusion model can capture general trends of fluctuations in energy and magnetism (related to the specific heat and susceptibility, respectively), GANs cannot describe such quantities. Importantly, the configurations produced by the diffusion model show power-law correlation between spins at the critical temperature, yet we find quantitative differences in the values of the correlation function at other temperatures when compared to those obtained from the Monte Carlo simulations. Our work emphasizes that while diffusion models are promising approaches for molecular ensemble generation, detailed analysis of various aspects of the generated ensemble is crucial for application to scientific research.

## 2. Methods and background

*2.1 Ising model*

We use an Ising model[41] defined on a square lattice with ferromagnetic interactions defined by the Hamiltonian $H = -J \sum_{i,j} s_i s_j$ where the sum runs over first nearest neighbors, $i$ and $j$, and the spin variables $s_i$ can be +1 or -1. The exchange variable, $J$, determines the strength of the interactions. This system exhibits a second order phase transition at the critical $T_c \sim 2.3 T^*$ from a paramagnetic phase to a ferromagnetic phase. The phase transition is characterized by a set of critical exponents that belong to a universality class that includes disparate systems.[26]

We generated Ising configurations using a MC algorithm introduced in Ref. [42]. We used a 64x64 lattice and extracted 1000 configurations for each temperature of interest. Each configuration to extract properties and to train the ML models is separated by 40960 spin flip attempts in the system to de-correlate configurations. Before the production runs, we thermalize the systems via 40,960,000 Monte Carlo move attempts. The code used to generate the Ising models for training the diffusion model is published on nanoHUB as an interactive tutorial.[43]

*2.2 Diffusion model*

The diffusion model consists of two processes: a forward diffusion in which noise is added to the original images ($x_0$) iteratively and a reverse generative process that reconstructs the original images by denoising the images in steps using an image-to-image neural network model named conditional UNET that will be discussed in section 2.3. We provide a brief description here and refer the interested reader to the manuscript by Ho et al.[17] for further details. Figure 1a depicts the schematics of a diffusion model. The forward diffusion process ($q$) is a fixed transformation that adds Gaussian noise ($\mathcal{N}$) to the ground-truth data ($x_0$) according to the following equation,

$$q(x_{1:N_t}|x_0) = \prod_{t=1}^{N_t} q(x_t|x_{t-1}) \qquad (1)$$

$$q(x_t|x_{t-1}) = \mathcal{N}(x_t; \sqrt{1-\beta_t}x_{t-1}, \beta_t I) \qquad (2)$$

Here, $\mathcal{N}(x_t; \sqrt{1-\beta_t}x_{t-1}, \beta_t I)$ represents a normal distribution whose output, mean, and variance are $x_t$, $\sqrt{1-\beta_t}x_{t-1}$, and $\beta_t I$, respectively. $N_t$ and $\beta_t$ are the number of diffusion steps and scheduled variances, respectively. We utilize linear scheduler of $\beta_t$ that varies from $\beta_1 = 10^{-4}$ to $\beta_{N_t} = 0.02$ and $T=1000$ as proposed by Ho et al.[17] Lastly, for an arbitrary step $t$, the forward step can be defined as,

$$x_t(x_0, \varepsilon) = \sqrt{\bar{\alpha}_t} x_0 + \sqrt{1-\bar{\alpha}_t}\varepsilon \qquad (3)$$

where $\varepsilon = \mathcal{N}(0, I)$, $\bar{\alpha}_t = \prod_{s=1}^{t} \alpha_s$, and $\alpha_s = 1 - \beta_s$.

The reverse generative process is defined as a learned Gaussian transition that denoises the latent distribution stepwise to obtain the target distribution by the following equations where $\mu_\theta$ and $\Sigma_\theta$ are the predicted mean and variances,

$$p_\theta(x_{0:N_t}) = p(x_{N_t}) \prod_{t=1}^{N_t} p_\theta(x_{t-1}|x_t, T) \qquad (4)$$

$$p_\theta(x_{t-1}|x_t, T) = \mathcal{N}(x_{t-1}; \mu_\theta(x_t, t, T), \Sigma_\theta(x_t, t, T)) \qquad (5)$$

Here $\theta$ and $T$ represent the parameters of the UNET and the temperature of the Ising system that is used as a conditional input to the UNET. $p(x_{N_t})$ is the initial image to be denoised and its values are set as $\mathcal{N}(0, I)$. $\Sigma_\theta$ is fixed as $\beta_t I$ and is not trainable, i.e. $\Sigma_\theta(x_t, t, T) = \Sigma(t)$. The mean of the distribution is $\mu_\theta(x_t, t, T) = 1/\sqrt{\alpha_t}(x_t - \frac{\beta_t}{\sqrt{1-\bar{\alpha}_t}}\varepsilon_\theta(x_t, t, T))$, where $\varepsilon_\theta(x_t, t, T)$ is the output of UNET with equal dimensions as $x_t$.

The objective of the diffusion model is to minimize the negative log likelihood of obtaining original image $x_0$ given the reverse process $p_\theta$, i.e. $\mathbb{E}[-\log(p_\theta(x_0))]$. Ho et al.[17] showed that this is variationally bound by

$$\mathbb{E}_q\left[D_{KL}\left(q(x_{N_t}|x_0)|p(x_{N_t})\right) + \sum_{t>1} D_{KL}(q(x_{t-1}|x_t,x_0)||p_\theta(x_{t-1}|x_t)) - \log p_\theta(x_0|x_1)\right] \quad (6)$$

In the implementation used, the above objective function is simplified as $L = \|\varepsilon - \varepsilon_\theta(x_t, t, T)\|^2$. Here, $\varepsilon_\theta(x_t, t, T)$ is the output of UNET for the reverse process whereas $\varepsilon$ is $\mathcal{N}(0, I)$ from the forward process. The UNET models for the diffusion process were trained for 300 epochs and the model with minimum loss was used to evaluate its performance on the test set.

**Training.** The diffusion model #1 (DiffIsing1) was trained using 1000 frames of MC simulations per temperature, for temperatures between $1.6\,T^*$ and $3.4T^*$ at intervals of $0.2\,T^*$. The diffusion model #2 (DiffIsing2) was trained using the same number of samples per temperature and temperature range, but the with a finer $T^*$ interval around the critical temperature: 0.05 for $2.0 \leq T^* \leq 2.5$. The diffusion models were tested for the range of $1.0 \leq T^* \leq 4.0$ with $T^*$ interval of 0.1 to evaluate its ability to extrapolate to temperature ranges not within the training set. The objective function for the diffusion model did not include energy, magnetism, specific heat, or magnetic susceptibility.

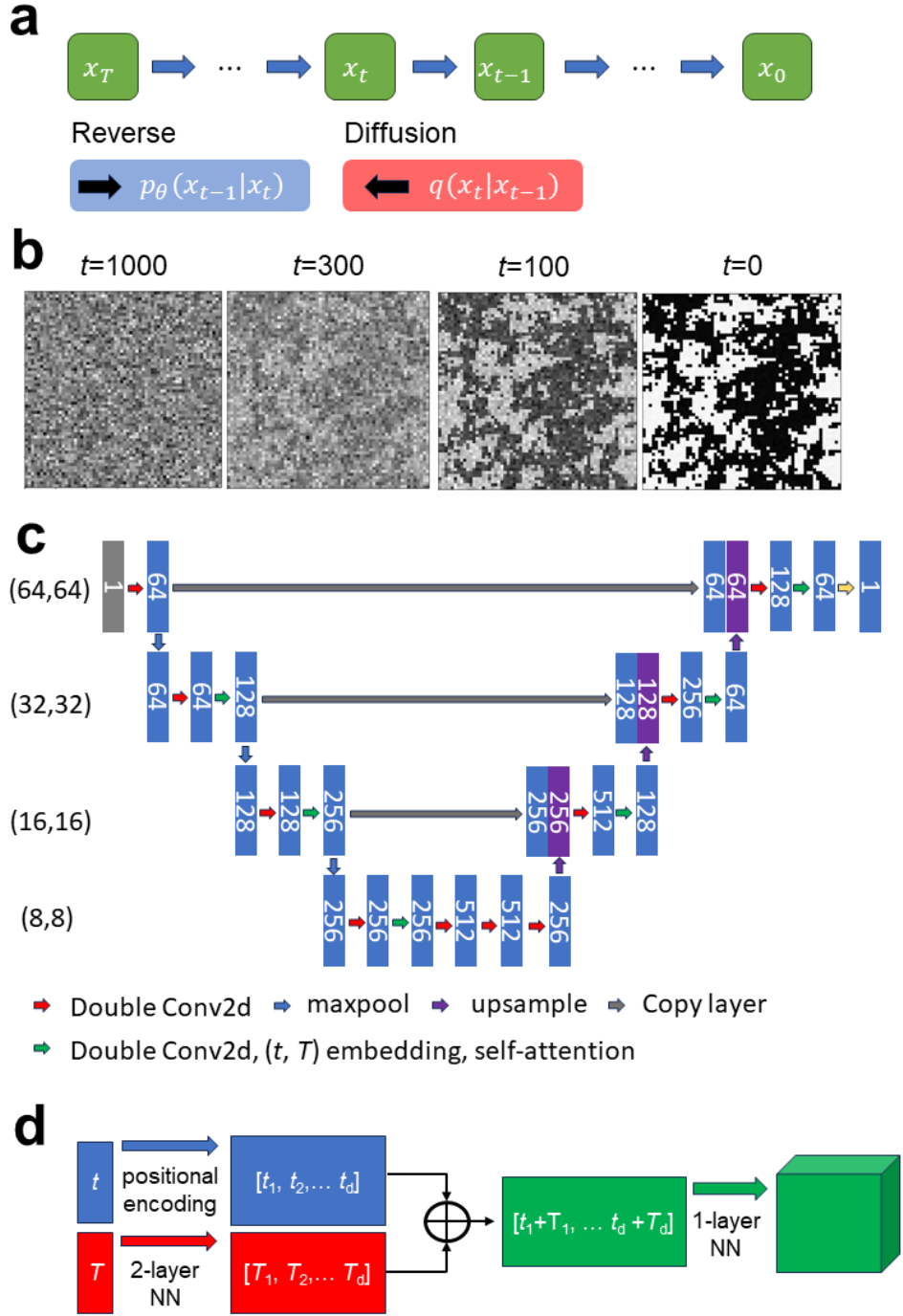

**Figure 1. Schematics of diffusion model.** (a) The reverse and forward (diffusion) process. (b) Time step dependent latent array of Ising. The initially random array ($t$=1000) is converted to Ising ensemble with the corresponding temperature through the reverse process ($t$=0). (c) UNET architecture. The numbers within blocks represent the channel dimensions while the numbers in parentheses represent the image dimensions. (d) Time and temperature embedding.

*2.3 UNET architecture*

The generative steps utilize a UNET architecture similar to PixelCNN++[44] with group normalization as suggested by Ho et al.[17], as depicted in Fig. 1c. The 64×64 images are down-sampled by four feature map resolutions (64 × 64 → 32 × 32 → 16 × 16 → 8 × 8). Each double Conv2d block consists of 2D convolution → group normalization → Gaussian Error Linear Unit (GELU) activation → 2D convolution → group normalization.

The diffusion time and Ising temperature embeddings are depicted in Fig. 1d. The diffusion time ($t$) embedding is calculated by transformer sinusoidal positional encoding ($PE$) proposed by Vaswani et al.[45]

$$PE_{(t,2i)} = \sin(t/10000^{2i/d}); PE_{(t,2i+1)} = \cos(t/10000^{2i/d}) \tag{7}$$

where $i$, $d$ are the current dimension and total dimension of encoding. The conditional variable, temperature of the Ising system, is subjected to two linear layers of dimensions ($1 \rightarrow d \rightarrow d$) followed by rectified linear unit activations to obtain temperature embedding. The sum of temperature and time embeddings is processed through Sigmoid linear unit and a linear layer then reshaped into the image dimension to be added to the image in each layer of UNET.

*2.4 Generative adversarial networks*

A generative adversarial network (GAN) is a generative model based on game theory. The model consists of two models, the generator ($G$) and the discriminator ($D$), trained simultaneously through an adversarial approach. The generator produces fake samples as a function of some random noise vector ($z$) that is sampled from a Gaussian distribution. For this study, we use a conditional generator whose input vector consist of $1 \times 200$ vector, with first 100 entries assigned as random Gaussian numbers and the rest are assigned by a temperature embedding obtained from two-layer neural network equivalent to that used for the diffusion model. Five transposed convolution neural networks followed by batch normalization and Rectified Linear Unit (ReLU) activation functions are used to convert the input array into the Ising ensemble of desired dimensions. The discriminator's objective is to distinguish between real data and fake data produced by the generator. We use a conditional discriminator with temperature embedding obtained from the same architecture as the generator. The discriminator architecture consists of 5 layers of convolution neural networks followed by batch normalization and activation functions. The activation functions for the first four layers were leaky ReLU and the function for the last layer was sigmoid. The real data were labeled as 1 while fake data were labelled as 0 in training the discriminator.

We utilize the Wasserstein GAN to reduce the effect of mode collapse. The training process involves a two-player zero-sum game where the generator produces fake data to fool the discriminator while the discriminator endeavors to distinguish between real and fake data. The objective function is:

$$\min_G \max_D V(D, G) = \mathbb{E}_{x \sim p_{\text{data}}(x)}[D(x|T)] - \mathbb{E}_{z \sim p_z(z)}[D(G(z))] \tag{8}$$

where $p_{\text{data}}(x)$ is the distribution of Ising data from MC and $p_z$ is the noise distribution. The first term on the right-hand side represents the expectation of discriminator's score over the Ising data sampled from MC results. The second term on the right-hand side represents the expectation of the discriminator's score over generated data. The discriminator is subject to the Lipschitz constraint through weight clipping.

## 3. Results and discussion

Our goal is to assess the ability of two generative models, diffusion and GAN, to predict Ising configurations over a range of temperatures. These configurations should follow the Boltzmann distribution and accurately describing the physics of the model requires capturing the entire distribution and not just average properties. In addition, the Ising model exhibits a second order phase transition and the configurations at the critical temperature ($T_c$) exhibit non-trivial properties. We start by comparing the average energy ($\langle U(T) \rangle$) and magnetic moment ($\langle M(T) \rangle$) as well as derived quantities associated with fluctuations specific heat ($C$), and magnetic susceptibility ($\chi$). These quantities are defined as[46]

$$\langle M(T) \rangle = \frac{1}{NL^2} \left| \sum_k^N \sum_i s_i^k \right| \tag{9}$$

$$\langle U(T) \rangle = -\frac{J}{N} \sum_k^N \sum_{\langle i,j \rangle} s_i^k s_j^k \tag{10}$$

$$C = \frac{\partial U}{\partial T} = \frac{1}{L^2 T^2} (\langle U^2 \rangle - \langle U \rangle^2) \tag{11}$$

$$\chi = \frac{1}{L^2 T} (\langle M^2 \rangle - \langle M \rangle^2) \tag{12}$$

Here, $s_i^k, L, N, T,$ and $\langle i,j \rangle$ are spin at lattice position $i$ and configuration $k$, length of the system ($L^2$ is the number of spins), number of frames, temperature, nearest neighbor pairs $i, j$, respectively. The average energy and magnetization are obtained as ensemble averages over the spin configurations generated using Monte Carlo simulations or the generative models. Due to the symmetry of the model with respect to flipping all the spins, we report the absolute value of the magnetization, Eq. 9, for the MC simulations. We note that below the critical temperature the system is not ergodic and the sign of the magnetization of the frames from MC are highly correlated as the probability of flipping the overall magnetization in the MC simulations is negligible. This is clearly seen in the distribution of magnetizations throughout our MC simulations, see Supplementary Materials (SM) Fig. S3. However, each frame from the ML models is generated independently and the signs of the frames are not correlated. This leads to bimodal peaks in the magnetization distribution (SM Fig. S3b) with the means around $\pm 1$ with small standard deviations. This is different from the bimodal peak at the critical temperature of $T_c$=2.3 where the means of the distributions are around 0. Therefore, each frame of the ML models for the temperatures below $T_c$=2.3 we report: $\frac{1}{NL^2}\sum_k^N |\sum_i s_i^k|$. The effect of this choice is discussed more in the SM section 1. The specific heat and magnetic susceptibility are obtained from the fluctuation expressions over the same configurations.

Figure 3 compares the four quantities of interest obtained with the MC model and the two diffusion models with the analytical solution. The shaded region shows the range of temperatures used for training the ML models. The values for these quantities corresponding to the analytical solution were obtained from Ref. [47] using WebPlotDigitizer.[48]

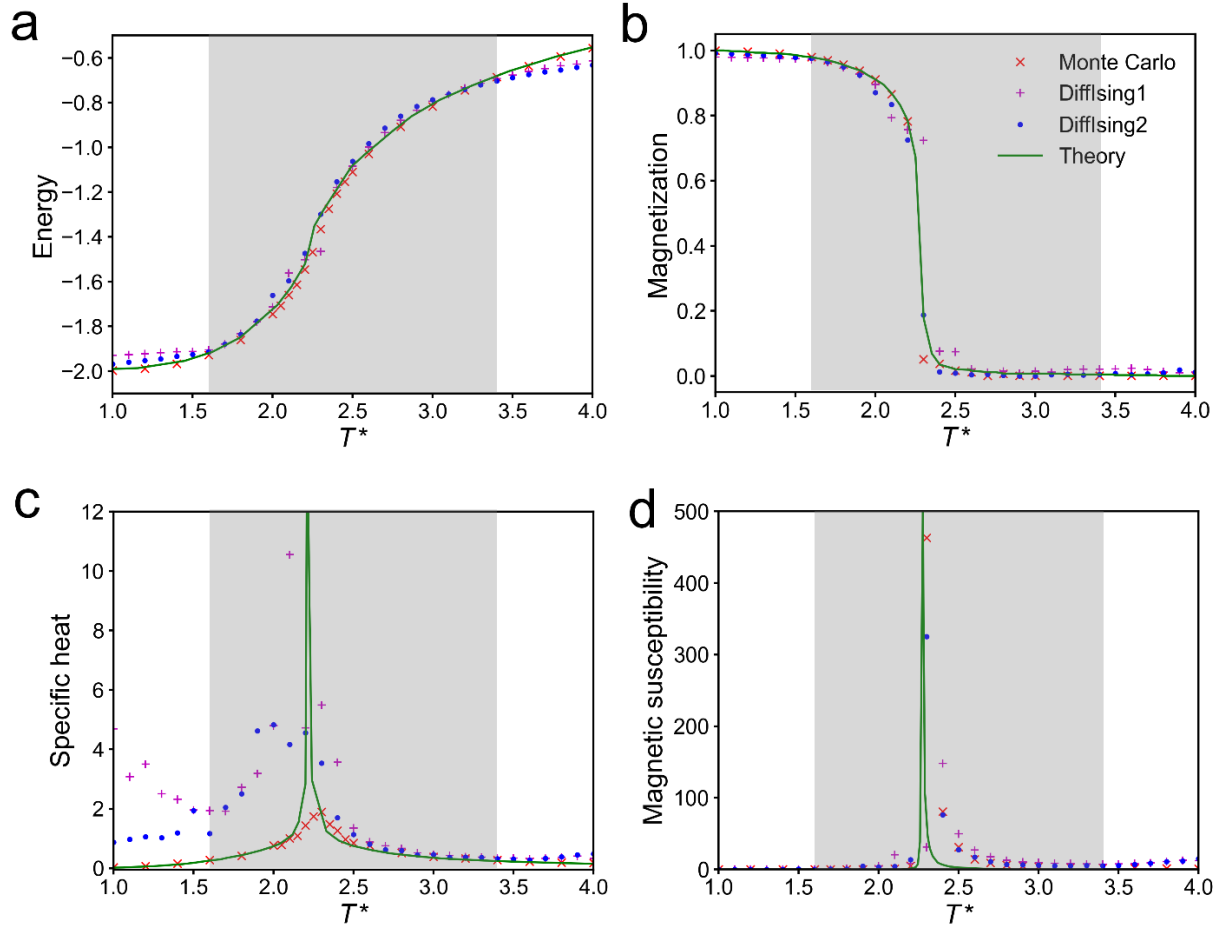

Figure 3. Comparison of Ising ensembles obtained from MC simulations, 2 generative diffusion models, and theory. (a) Energy per spin, (b) magnetization per spin, (c) specific heat, (d) magnetic susceptibility. Theoretical values are obtained following Ref. [47]. The temperature range used for training is shaded in gray.

As expected, the MC simulations follow the analytical result quite closely, the deviations near the critical temperature are due to small size effects and the divergence in correlation length at the critical point. Interestingly, both diffusion models describe the energy and magnetization as a function of temperature rather accurately, including some ability to extrapolate outside the range of training. Minor discrepancies can be observed for DiffIsing1 near $T_c$ and at the extreme ends of the temperature range. Near $T_c$, the diffusion model displays correct values of the energies and magnetism, but the energies and magnetism change linearly for $2.1 \leq T^* \leq 2.3$ then abruptly changes at $T^*=2.4$ (see Fig. S5 for log scale plots of the data). This inaccuracy in describing the rate of change is because the training set contained a single point ($T^*=2.2$) in that range ($T^*=2.1$, 2.2, and 2.3). Such results indicate that the training dataset should be chosen carefully as the thermodynamic behavior at this range of temperatures is more rapidly changing than at other temperatures.

Interestingly, the diffusion models can reproduce the large fluctuations around the critical temperature, see Figs. 3(c) and (d). Both models describe the magnetic susceptibly quite accurately, but display erroneously large values of the specific heat for $T<T_c$. This can be explained by analyzing the distribution of energies from each frame for the MC and the diffusion models, shown in Fig. S4. For MC (Fig. S4a), the standard deviation (STD) of the energies is 0.002-0.005 for $T=1.0$ and 1.2 with energy per spin nearly equal to -2. On the other hand, the energies of diffusion models for these temperatures are around -1.95 with STD in the range 0.01-0.04. While this difference may be small, the specific heat expression has a $1/T^2$ term that amplifies errors in the low-temperature fluctuations. Therefore, for the low temperature regime, the broader distributions from the generative models lead to significant values of the specific heat. This difference may also be affected by the difference in the sampling method of MC and the diffusion models. In the non-ergodic regime of $T<T_c$, the ensembles from MC represent the sampling from a single initial state while all of the frames from the generative models are from independent states.

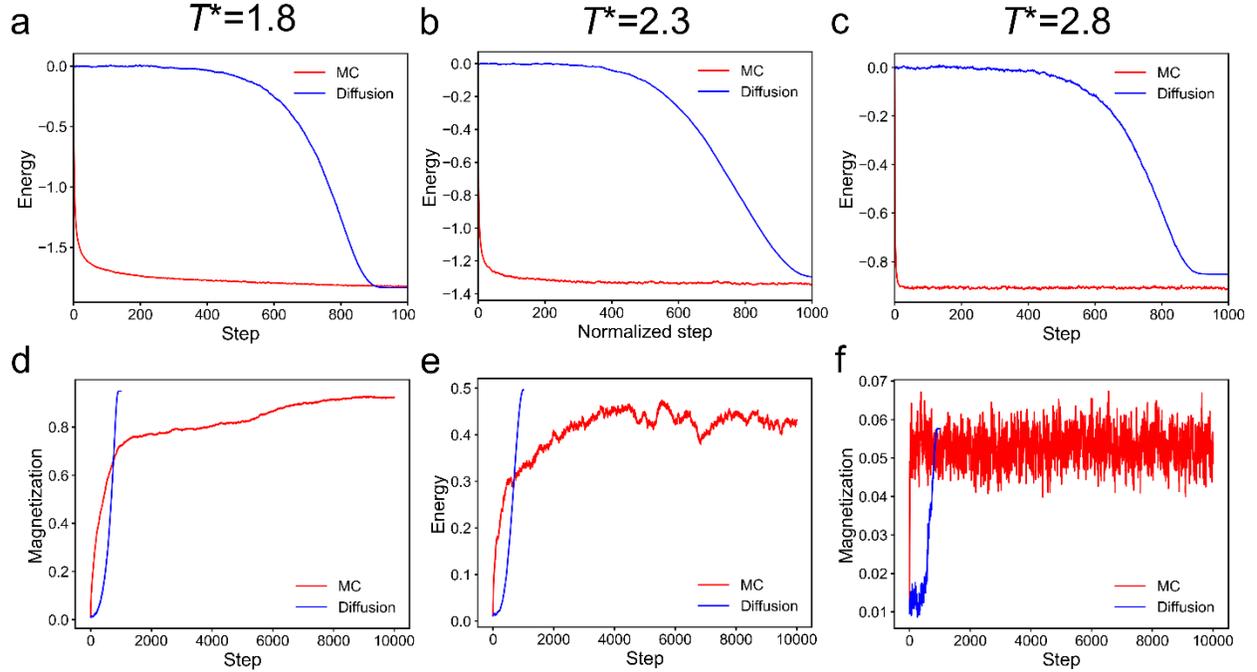

Figure 4. Evolution of energy (a-c) and magnetization (d-f) for Monte Carlo and DiffIsing2 for temperatures of (a, d) 1.8, (b, e) 2.3, and (c, f) 2.8. Step corresponds to the diffusion step for diffusion and $L^2$ trials movements for the MC.

To better understand how the diffusion model generates the Ising configurations from the initial random configuration, we follow the evolution of the energy and magnetization as a function of the generative steps. We compare the DiffIsing2 model (blue lines in Figure 4) with the evolution towards equilibrium followed by the Metropolis algorithm (red lines), also starting from a random configuration. While the Metropolis approach is not designed to capture how systems approach equilibrium, we believe the comparison is informative. The results in Fig. 4 represent the averages over 100 independent initial random samples. Each normalized step represents $L^2$ ($64^2$) trial moves for MC and one diffusion step for diffusion. Interestingly, while the MC simulation starts moving towards equilibrium immediately, the diffusion model exhibits an induction stage, where the energy and magnetization do not change significantly. We speculate that this may be because MC evolves the system to minimize free energy while diffusion uses a process of gradual removal of the noise from a random system of $(p(x_{1000}))$ towards a target system $(p(x_0))$ without explicit consideration of the thermodynamic quantities of the system.

Such results demonstrate the difference between MC trial moves and diffusion steps. The trial moves of MC are sequential and local transformations. Therefore, the system can sample local energy minima and take many trials to overcome the energy barriers at low temperatures. On the other hand, the diffusion steps transform the initial random array into a targeted distribution in a global and collective manner. These steps are not affected by local energy barriers. In addition, while temperature is a conditional input to our diffusion model, it provides information on the final distribution and does not necessarily affect the magnitude of the transformations from diffusion steps. Therefore, while the generative model may lead to the same thermodynamic quantities as those obtained from physics-based simulations, the equilibration process is fundamentally different.

We also trained generative adversarial network (GAN) models for the Ising system. The GAN models were trained with the same data as DiffIsing2. The results (Fig. 4) show that conditional GAN models with the temperature as input condition cannot capture the physics of the Ising system. Therefore, we added the mean squared error of the magnetization of the generated ensemble as an additional loss term to the generator, $L = \frac{1}{N}\sum_1^N (\bar{s}_{GAN} - \bar{s}_{MC})^2$. The resulting model will be denoted as GAN, $M \subset L$ to indicate that the model utilized magnetization as a loss function during its training. Such change increased the accuracy for the energy and magnetism for the temperatures within the training set but was not sufficient to capture the specific heat and magnetic susceptibility.

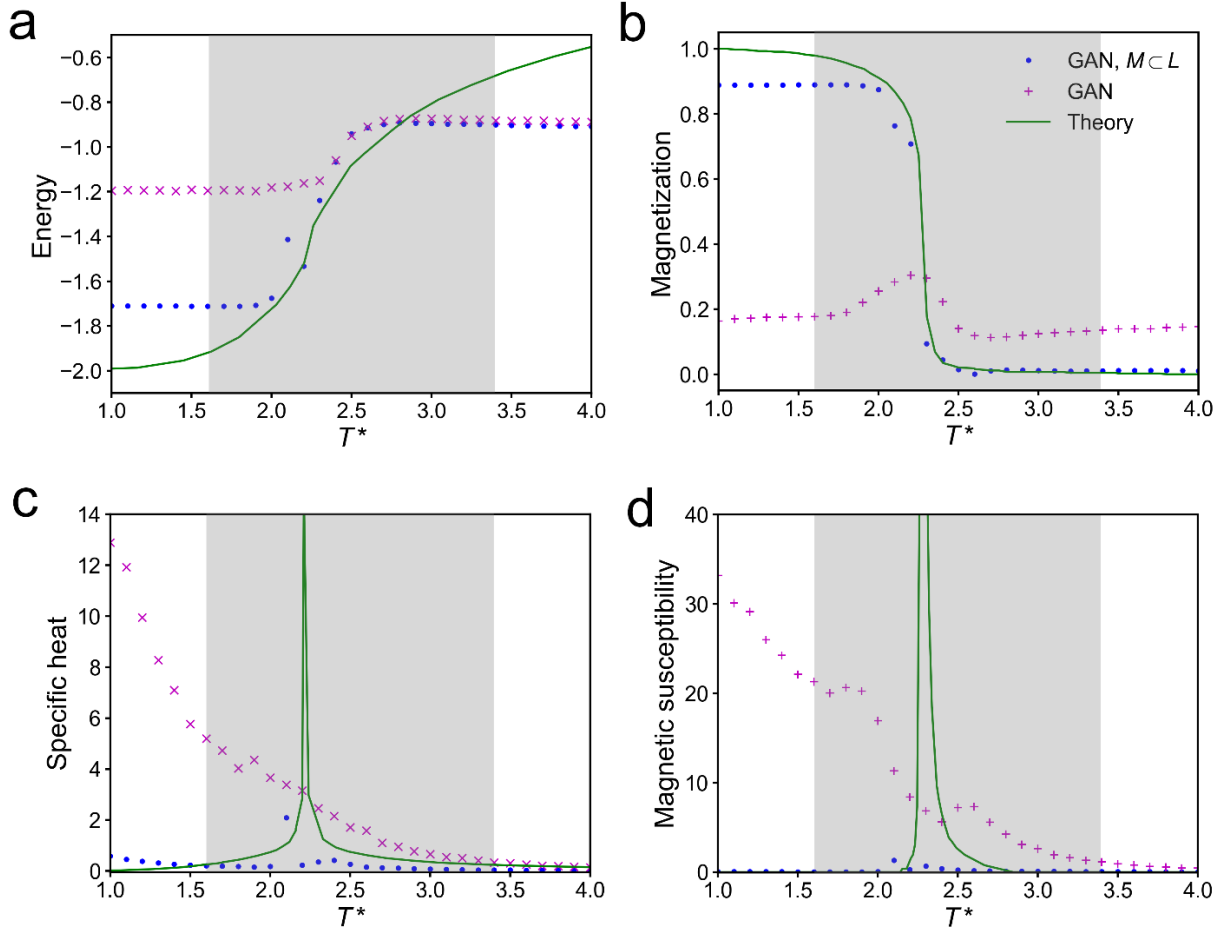

Figure 4. Comparison of Ising ensembles obtained from generative adversarial networks with and without magnetization as part of loss function, and theory. (a) Energy, (b) magnetization, (c) specific heat, (d) magnetic susceptibility. Temperature range used for training is shaded in gray.

In addition to the thermodynamic quantities, we are interested in characterizing the details of the configurations, especially those around the critical point. To this end, we calculated the spin-spin correlation function $G$ and the correlation length ($\xi$) defined as

$$G(r_{ij}) = \langle s_i s_j \rangle - \langle s_i \rangle \langle s_j \rangle \tag{13}$$

where $r_{ij}$ is the distance between spins $i$ and $j$ taking the lattice parameter of be unity. The correlation function can be empirically shown to be described by $G(r) = r^{-\eta} e^{-\xi r}$, where $\eta$ is a critical exponent and $\xi$ the correlation length. The divergence of $\xi$ and a power law correlation function at the critical temperature indicate scale invariance and are central to second-order phase transitions. Importantly, the correlation is a self-similar power law at $T_c$.

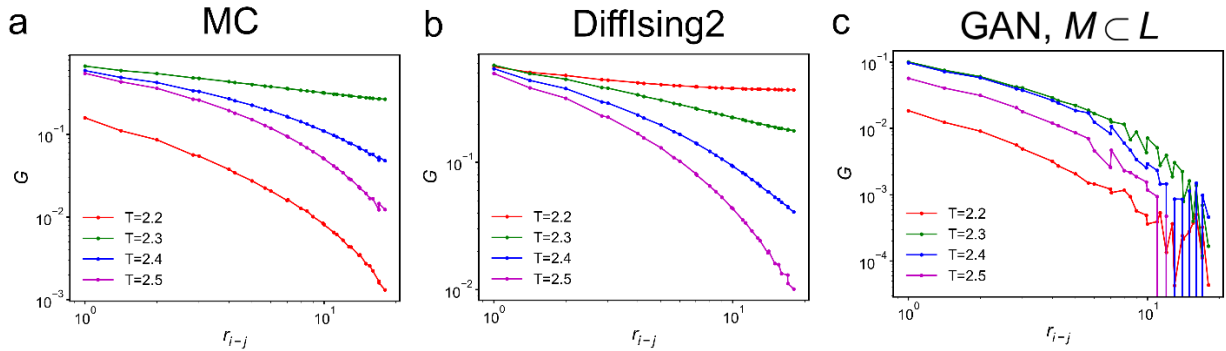

Figure 5. $G$ as function of lattice distance for (a) Monte Carlo, (b) DiffIsing2, and (c) GAN, $M \subset L$.

The MC results exhibit the divergence of $\xi$ and scale invariance at $T^*=2.3$, with the critical exponent $\eta$ of 0.308 (see Fig. S4 for plot of the critical temperature result) in good agreement with previous simulation results.[49] For the diffusion models, we observe similar behavior of divergence at $T^*=2.3$, but with critical exponent of 0.398. This indicates that the diffusion model is capable of qualitatively capturing criticality although replicating the behavior in a quantitatively accurate manner is more difficult. In addition, we also note that the analytical value of $\eta$ for infinite 2D Ising system is 0.25,[50] but the values from finite system simulations can depend on the simulation conditions and fitting schemes, leading to a range of reported values from 0.138[51] to 0.4.[52] In addition, we observe that the values of $G$ are larger at $T^*=2.2$ and display power-law behavior unlike the results from Monte Carlo simulations. Lastly the GAN model results depicted in Fig. 5c demonstrate that the model is incapable of capturing the long-range divergence of correlation length.

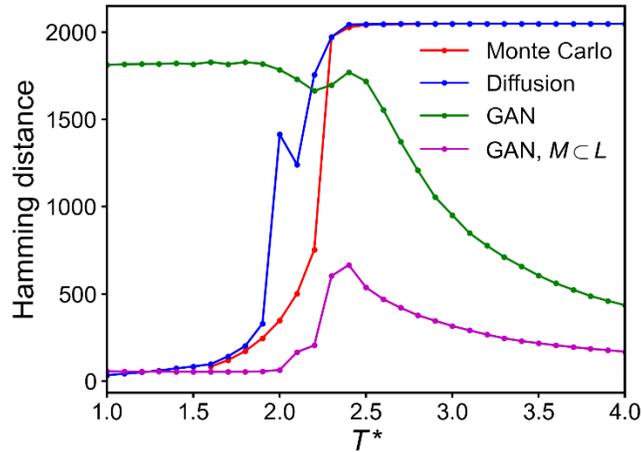

Figure 6. Hamming distance versus temperature for Ising ensembles obtained from Monte Carlo, diffusion model #2, and GAN models.

To further analyze the performance of generative models, we analyze the diversity of the obtained ensemble configurations by calculating the average hamming distance as shown in Fig. 6. For the MC and DiffIsing2 model, we observe high similarity between configurations at low temperatures and high diversity above $T_c$ as noted by the values of the Hamming distances. The trend as well as numerical values are similar for the high and low temperature regions with minor variation around $T_c$. On the other hand, GAN ($M \subset L$) display generally low hamming distance, especially at high temperatures, indicating that the outputs of the generator were more limited than the original data set. Furthermore, for GAN without ($M \subset L$), we see higher hamming distance at low temperatures, followed by decrease in the hamming distances as the temperature is increased. These results again indicate that our GAN model is not correctly describing the Ising data distribution, an analysis that is not evident from the energy and magnetization in Fig. 2 or via visual inspection as shown by snapshots in Fig. 7.

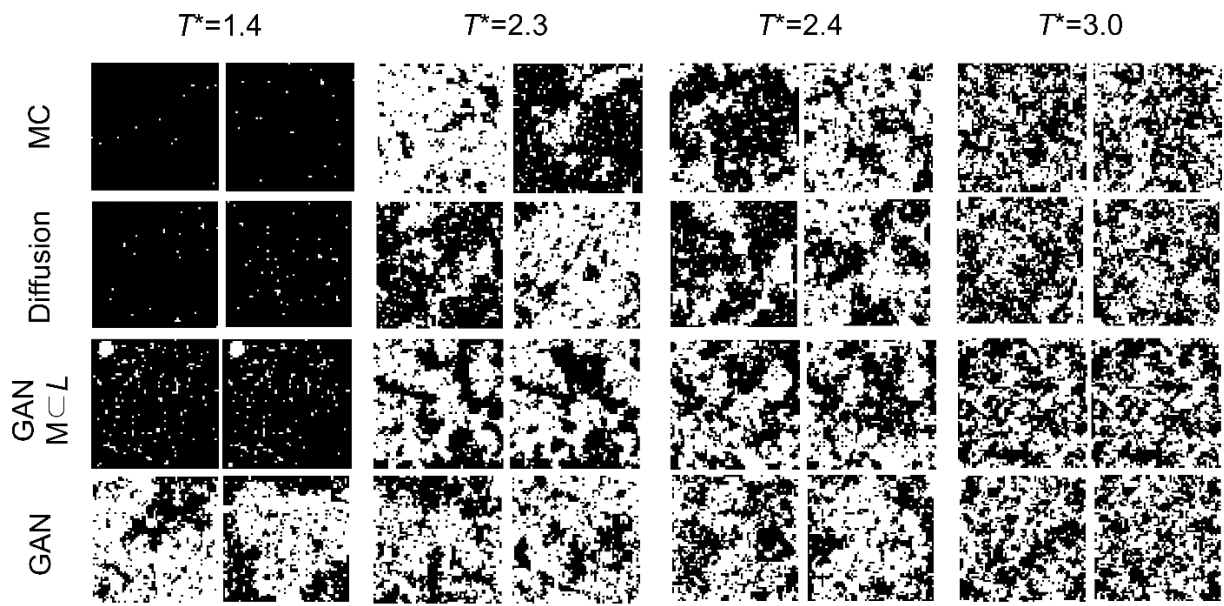

Figure 7. Snapshots of Ising ensembles at $T^*$ of 1.4, 2.3, 2.4 and 3.0 for Monte Carlo, DiffIsing2, GAN ($M \subset L$), and GAN. Two snapshots are depicted for each case.

## 4. Conclusions

In this study, we examined the effectiveness of generative machine learning models for Ising system. Diffusion models can capture thermodynamic quantities and their fluctuations, underscoring its potential as a robust tool in simulating physical systems. In addition, the model captures power-law correlations near the critical temperature. A critical element in achieving accuracy for diffusion model was the judicious selection of the training set, which necessitated a priori physics knowledge. In contrast, GAN required loss functions specifically targeting the thermodynamic quantities to partially capture the physics of the system. Furthermore, GANs exhibited a susceptibility to mode-collapse behavior, which can potentially undermine the diversity and reliability of generated ensembles, thereby limiting their applicability in exploring the comprehensive phase space of physical systems.

This study illuminates the potential and challenges of employing generative models in the scientific domain. While the diffusion model demonstrated accurate portrayal of the Ising model, its effectiveness is notably enhanced by the integration of domain-specific knowledge in the model training. GAN captured the physics of Ising model only partially, and understanding the areas of failure necessitated careful analysis of the generated results. In addition, while the generative models may not fully replace simulations, hybrid models utilizing generative models for structure generation that are fine-tuned with subsequent physics-based simulations may be developed to accelerate scientific investigations. Overall, the results show that domain expertise is crucial in training and analyzing machine learning models for scientific applications.

## Code availability

The codes associated with this work is available on:

https://github.itap.purdue.edu/StrachanGroup/generative_diffusion_Ising

## Acknowledgments

The authors acknowledge the support of Purdue University. B.R. and A.S. were supported by SCALE: U.S. Department of Defense Contract No. W52P1J-22-9-3009. B.H.L. was supported by National Science Foundation under DUE ATE #2000281, the Micro Nano Technology

## Data availability

The data associated with this work is available on:

https://github.itap.purdue.edu/StrachanGroup/generative_diffusion_Ising

## Author declarations

The authors have no conflicts to disclose.

[45] A. Vaswani, N. Shazeer, N. Parmar, J. Uszkoreit, L. Jones, A.N. Gomez, Ł. Kaiser, and I. Polosukhin, "Attention Is All You Need," *Advances in Neural Information Processing Systems*, 30 (2017).

[46] M.E. Tuckerman, *Statistical Mechanics: Theory and Molecular Simulation* (Oxford University Press, 2023).

[47] A. Codello, "Chapter 5: Ising model and phase transitions," Corpus ID: 131767819 (2013).

[48] Ankit Rohatgi, "WebPlotDigitizer," https://Automeris.Io/WebPlotDigitizer.

[49] P.H. Lundow, and I.A. Campbell, "Bimodal Ising spin glass in two dimensions: The anomalous dimension η" *Physical Review B* **97**(2), 024203 (2018).

[50] H.G. Katzgraber, and L.W. Lee, "Correlation length of the two-dimensional Ising spin glass with bimodal interactions," *Physical Review B* **71**(13), 134404 (2005).

[51] I. Morgenstern, and K. Binder, "Magnetic correlations in two-dimensional spin-glasses," *Physical Review B* **22**(1), 288–303 (1980).

[52] V.S. Dotsenko, and V.S. Dotsenko, "Critical behaviour of the phase transition in the 2D Ising Model with impurities," *Advances in Physics* **32**(2), 129–172 (1983).

[53] H.G. Katzgraber, and L.W. Lee, "Correlation length of the two-dimensional Ising spin glass with bimodal interactions," *Physical Review B* **71**(13), 134404 (2005).

[54] I. Morgenstern, and K. Binder, "Magnetic correlations in two-dimensional spin-glasses," *Physical Review B* **22**(1), 288–303 (1980).
26

# Supplementary Materials for Thermodynamic Fidelity of Generative Models for Ising System

Brian H. Lee, Kat Nykiel, Ava E. Hallberg, Brice Rider, Alejandro Strachan[a]

Author affiliations
*School of Materials Engineering and Birck Nanotechnology Center, Purdue University, West Lafayette, Indiana 47907, USA*

Author email
[a] Author to whom correspondence should be addressed: strachan@purdue.edu

# 1. Magnetization analysis

We chose two methods of analyzing the ensemble-averaged magnetization ($\langle M \rangle$). In the first method, we calculated the sum of spin values of the entire trajectory and took the absolute value of the sum as given by $\langle M \rangle = \frac{1}{NL^2}\left|\sum_k^N \sum_i s_i^k\right|$. In the second method, we used the absolute value of the sum of spins per frame as given by $\langle M \rangle = \frac{1}{NL^2}\sum_k^N\left|\sum_i s_i^k\right|$. The results for Monte Carlo (MC) simulations and generative model (DiffIsing2) are depicted on Fig. 1.

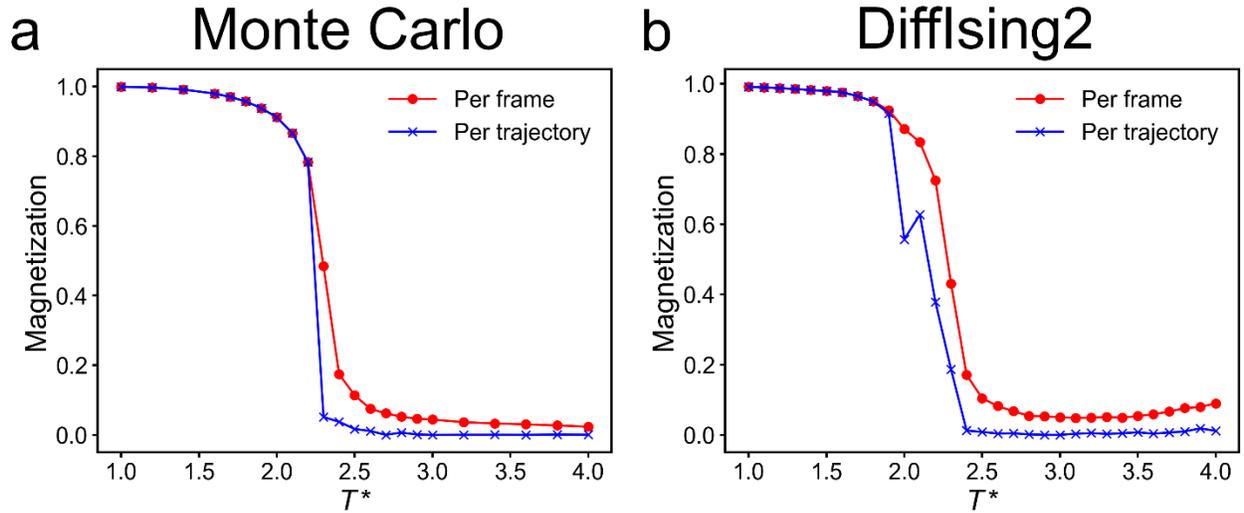

**Supplementary Figure 1.** $\langle M \rangle$ for MC (a) and DiffIsing2 (b). Red lines correspond to the $\langle M \rangle$ measured with absolute values applied per frame while blue lines are for those with absolute values applied for the entire dataset.

The results demonstrate that there can be significant differences in the $\langle M \rangle$ for the generative model in the non-ergodic regime ($T<T_c$ of 2.3). This is because the trajectories are not obtained in equivalent manner for MC and the generative models. For MC, the trajectories are obtained from one state that evolves through MC steps. For the non-ergodic regime, the states remain near the first local energy minima that they find, which can be majority spin up or spin down depending on the random number generator. On the other hand, each frame of the diffusion model is from an independent sample with varying initial states.

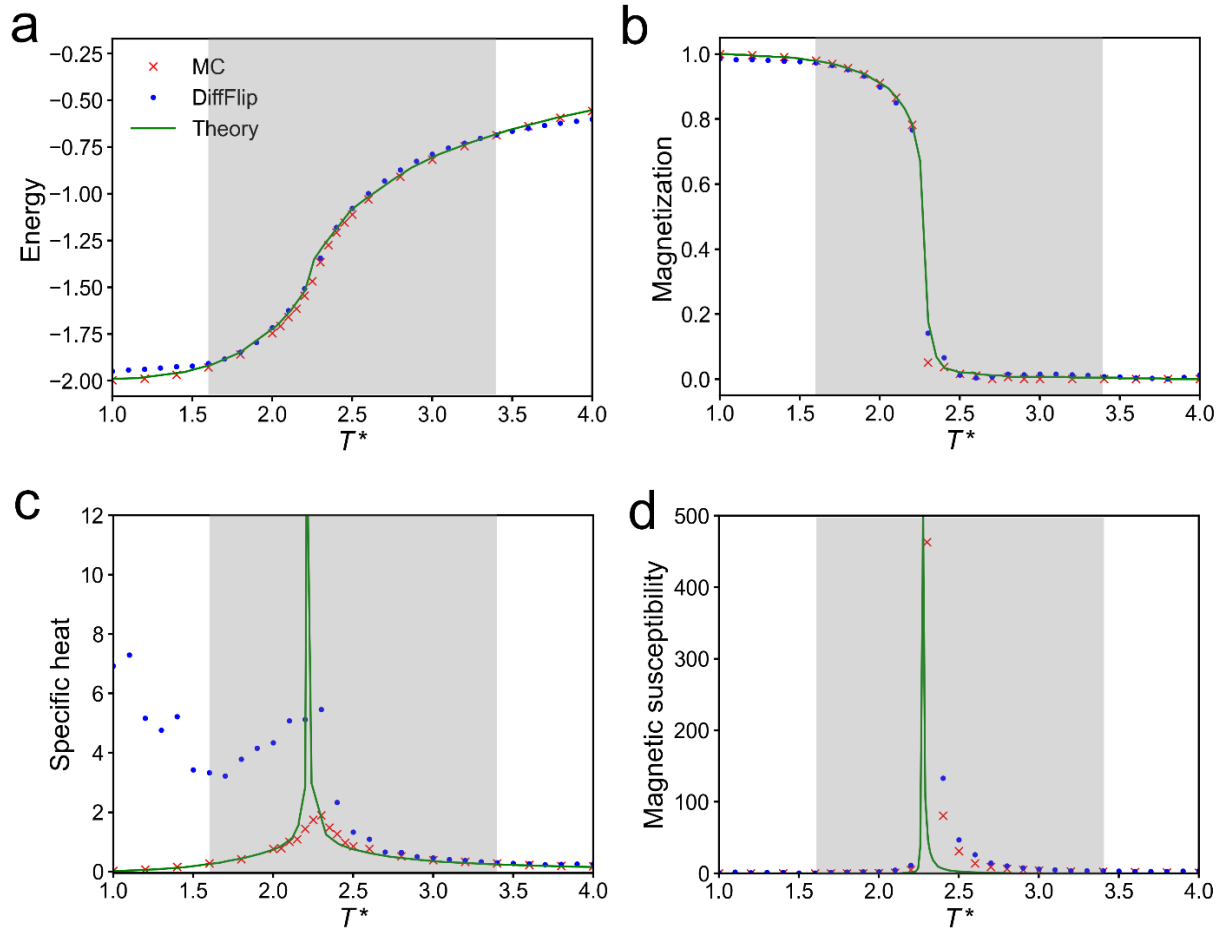

**Supplementary Figure 2.** Thermodynamic quantities from DiffFlip obtained from the dataset containing original MC data and those with the signs of the spins flipped. (a) Energy per spin, (b) magnetization per spin, (c) specific heat, (d) magnetic susceptibility. Theoretical values are obtained following Ref. [47]. The temperature range used for training ($1.6 \leq T \leq 3.4$) is shaded in gray.

As an additional analysis, we trained the diffusion model on the Ising ensembles from the original MC data and those with the signs flipped. We will denote the diffusion model as DiffFlip. In comparison with the results shown in Fig. 3 of the main manuscript, the results shown in Fig. S2 demonstrate that these quantities remain similar.

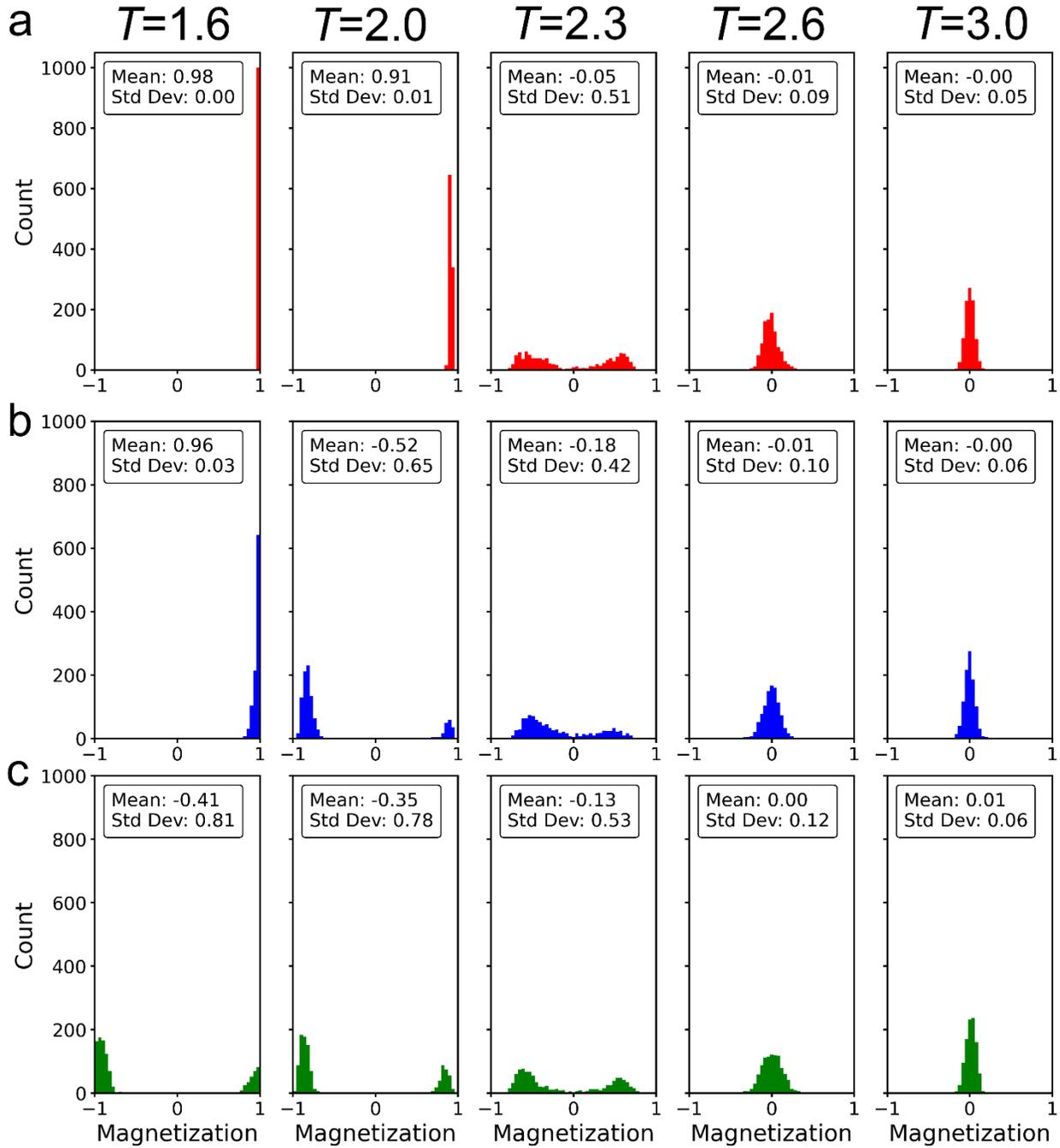

**Supplementary Figure 3.** Distribution of $M$ obtained from (a) MC, (b) DiffIsing2, (c) DiffFlip. Temperatures are 1.6, 2.0, 2.3, 2.6, and 3.0.

Lastly, we plot the distribution of $M$ for temperatures of 1.6, 2.0, 2.3, 2.6 and 3.0 on Fig. 2. The histograms represent the $M$ per frame for each temperature. For the critical temperature of $T^*=2.3$ and ergodic regime of $T^*=2.6$, the distributions of the MC (Fig. 3a), DiffIsing2 (Fig. 3b), and DiffFlip (Fig. 3c) are similar. However, for the non-ergodic case of $T^*$=1.6 and 2.0, the MC displays almost all frames with $M \sim 1$ while the diffusion models show bimodal distribution with means around $M \sim \pm 1$. While the two distributions are thermodynamically equivalent, the resulting $\langle M \rangle$ can be numerically different. Therefore, we take the absolute value of the entire trajectory for the MC simulations while the absolute values per frame are used for the generative models in the non-ergodic regime ($T^*$<2.3).

# 2. Additional data

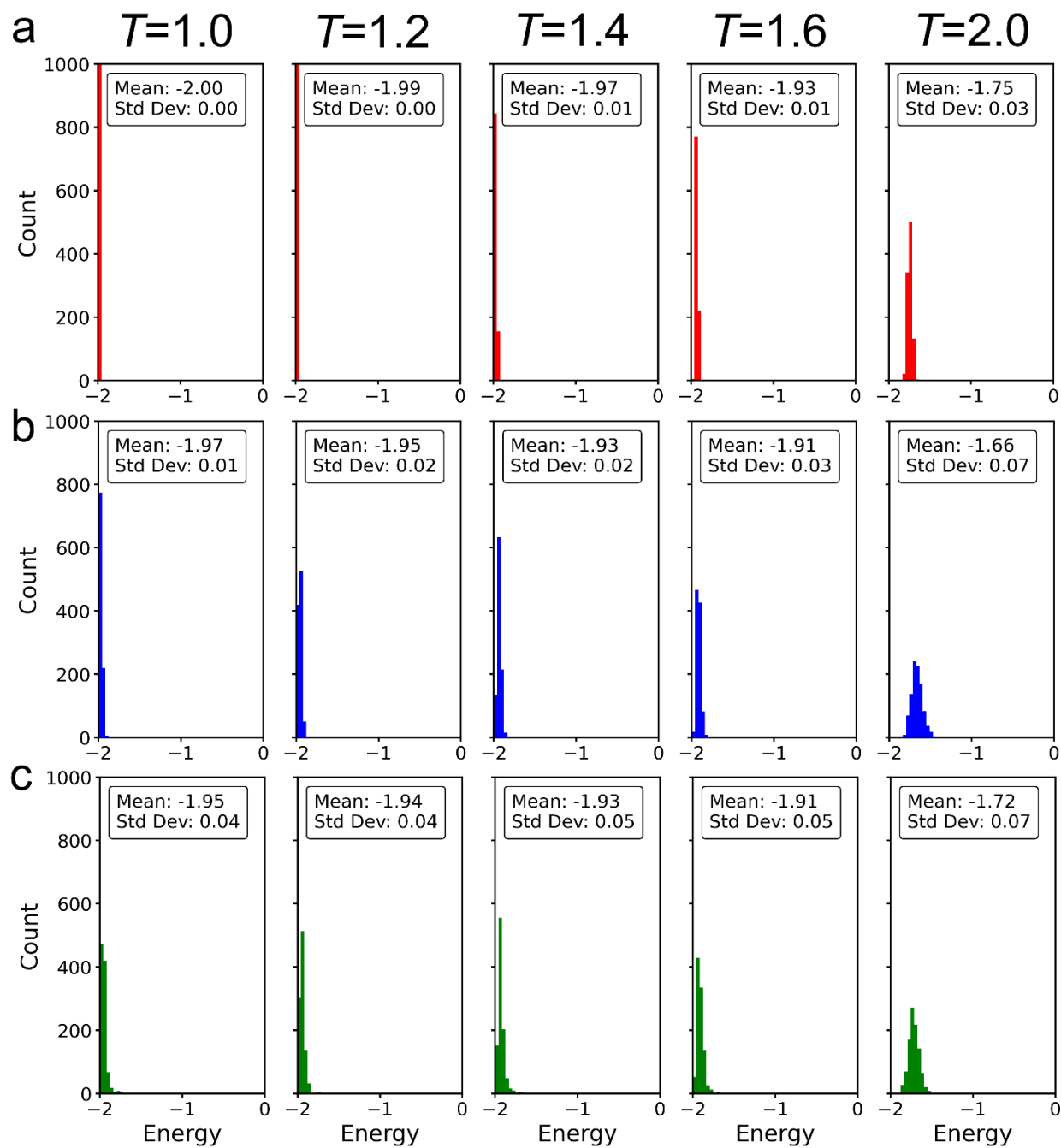

**Supplementary Figure 4.** Distribution of potential energies for (a) Monte Carlo, (b) DiffIsing2, and (c) DiffFlip models.

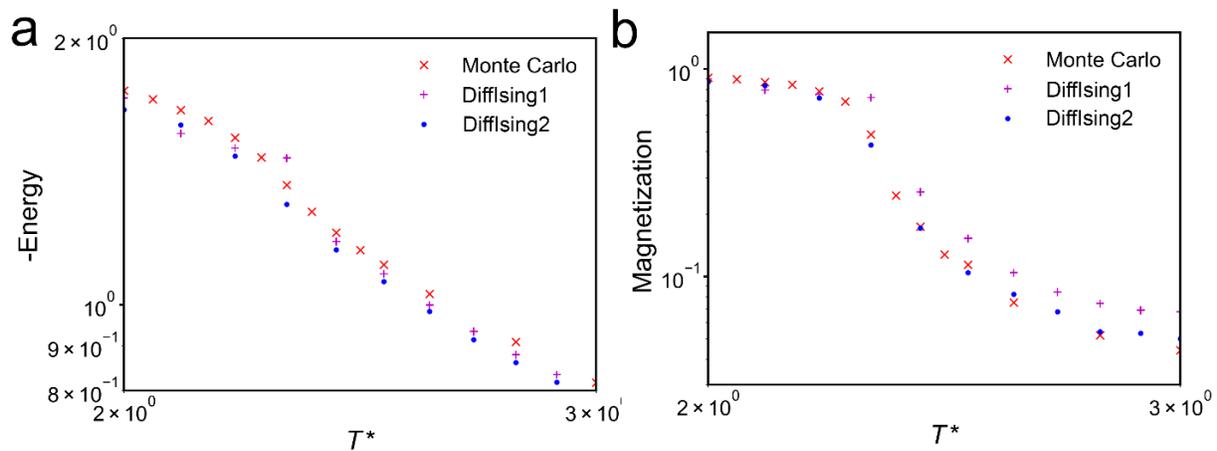

**Supplementary Figure 5.** Log-log plots of the energy and magnetization of the Monte Carlo, DiffIsing1, and DiffIsing2 models.

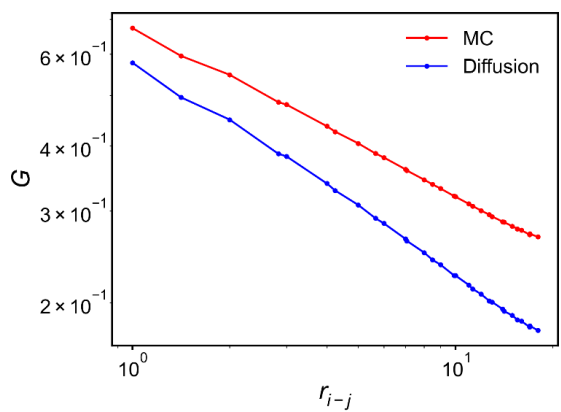

**Supplementary Figure 6.** Spin-spin correlation function at critical temperature T*=2.3 for the Monte Carlo and DiffIsing2 model.

33